\documentclass[preprint2]{aastex}
\received{2014 August 27}
\accepted{2014 September 13}

\slugcomment{ApJ}
\shorttitle{Revisiting Jovian-Resonance Induced Chondrule Formation}
\shortauthors{NAGASAWA ET AL.}

\begin{document}
\title{Revisiting Jovian-Resonance Induced Chondrule Formation}
\author{M. Nagasawa\altaffilmark{1},  K. K. Tanaka\altaffilmark{2}, 
H. Tanaka\altaffilmark{2}, T. Nakamoto\altaffilmark{3}, 
H. Miura\altaffilmark{4}, \& T. Yamamoto\altaffilmark{5}} 

\altaffiltext{1}{Interactive Research Center of Science, Tokyo Institute of Technology, 
2-12-1 Ookayama, Meguro-ku, Tokyo 152-8551, Japan}
\altaffiltext{2}{Institute of Low Temperature Science, Hokkaido University,
 Kita-19, Nishi-8, Kita-ku, Sapporo 060-0819, Japan}
\altaffiltext{3}{Department of Earth and Planetary Sciences, Tokyo Institute of Technology, 
2-12-1 Ookayama, Meguro-ku, Tokyo 152-8551, Japan}
\altaffiltext{4}{Graduate School of Natural Sciences, Nagoya City University, 1 Yamanohata, Mizuho-cho, Mizuho-ku, Nagoya 467-8501, Japan}
\altaffiltext{5}{Center for Planetary Science, Kobe University, 
7-1-48 Minamimachi, Minatojima, Chuo-ku, Kobe 650-0047, Japan}

\email{nagasawa.m.ad@m.titech.ac.jp}

\begin{abstract}
It is proposed that planetesimals perturbed by Jovian mean-motion resonances are the source of shock waves that form chondrules.  
It is considered that this shock-induced chondrule formation requires the velocity of the planetesimal relative to the gas disk to be on the order of $\ga 7~ {\rm km s^{-1} }$ at 1 AU.  
In previous studies on planetesimal excitation, the effects of Jovian mean-motion resonance together with the gas drag were investigated, but the velocities obtained were at most 8 ${\rm km s^{-1}}$ in the asteroid belt, which is insufficient to account for the ubiquitous existence of chondrules.  
In this paper, we reexamine the effect of Jovian resonances and take into account the secular resonance in the asteroid belt caused by the gravity of the gas disk. 
We find that the velocities relative to the gas disk of planetesimals a few hundred kilometers in size exceed 12 ${\rm km s^{-1}}$, and that this is achieved around the 3:1 mean-motion resonance.
The heating region is restricted to a relatively narrow band between 1.5 AU and 3.5 AU.  
Our results suggest that chondrules were produced effectively in the asteroid region after Jovian formation.  
We also find that many planetesimals are scattered far beyond Neptune. 
Our findings can explain the presence of crystalline silicate in comets if the scattered planetesimals include silicate dust processed by shock heating.
\end{abstract}

\section{INTRODUCTION}
\label{sec:intro}

{Chondrites reserve information about the early stage of 
the solar system that had been lost in the planets themselves.}  
Spherical chondrules, grains 0.1 mm to 1 mm in size composed of the silicates found in meteorites, are considered to have been formed from precursor particles that were heated and melted in flash heating events. 
These then cooled and resolidified in a short period of time ($\sim$ hours) in a protoplanetary disk (e.g., Jones et al. 2000). 
It is believed that the efficiency of chondrule formation was high because chondrules are major constituents in chondritic meteorites.  
So far, various mechanisms for chondrule formation have been proposed; however, it has not yet been concluded which process is predominant.  One mechanism for chondrule formation that has received considerable attention is heating by shock waves (Hood \& Horanyi 1991, 1993; Boss 1996; Jones et al. 2000).  
The heating process of chondrule precursors by shock waves has been investigated in detail in previous studies. It has been shown that the shock-wave heating model satisfies various constraints related to chondrule formation such as the peak temperature ($\sim 2000$ K) and short cooling time 
(Hood \& Horanyi 1991, 1993; Iida et al. 2001; Ciesla \& Hood 2002; Miura, Nakamoto, \& Susa 2002; Desch \& Connolly 2002; Miura \& Nakamoto 2005, 2006).

As plausible sites for such shock waves to occur, highly eccentric planetesimals have been proposed (Hood 1998; Weidenschilling, Marzari, \& Hood 1998).  
If the relative velocity between a planetesimal and the gas disk exceeds the speed of sound of the gas, a bow shock is produced.  
Weidenschilling et al. (1998) suggested that Jovian mean-motion resonances excite planetesimals. 
The formation of Jupiter in the gas disk induces resonances and strongly affects the motion of planetesimals around the asteroid belt ($\sim$ 2~AU--5~AU).  
The evolution of such planetesimals under the influence of gas drag has also been studied in detail (Ida and Lin 1996; Marzari et al. 1997; Weidenschilling et al. 1998; Marzari \& Weidenschilling 2002).  
The works showed that planetesimals migrate toward the sun due to gas drag even if their radii are on the order of 1000 km. 
During the migration from 4~AU to 3~AU, the eccentricities of the planetesimals are excited by multiple Jovian mean-motion resonances. 
The excited planetesimals can acquire further eccentricity up to about $e\sim 0.4$ ($\sim 6 {\rm km s^{-1}}$) during the trapping in the 2:1 resonance ($\sim$~3.3AU), provided Jupiter has an eccentricity larger than $e\ga 0.03$.
The excitation of eccentricity increases the gas drag, and the eccentricity and semi-major axis are quickly damped as a result. 
The orbits of many planetesimals are circular before they reach 2~AU. 

Marzari \& Weidenschilling (2002) showed that the velocity of 100 km--300 km-sized planetesimals relative to the gas disk reaches a maximum of 8 ${\rm km s^{-1}}$ ($e\sim$ 0.6 at 2~AU). 
Planetesimals with such high velocities are rare and the period in which these velocities are achieved is limited to on the order of $10^4$ yr.
Simulations of gravitationally interacting planetesimals suggest that the process is not very efficient, i.e., the area swept by chondrule-forming shocks over a period of 1 Myr--2 Myr is just $\la$ 1\% and the planetesimals need to be about half the size of the Moon to accumulate the speed required for chondrule formation 
(Hood \& Weidenschilling 2012).

\begin{figure*}\begin{center}
\includegraphics[width=13cm, bb=10 0 500 240]{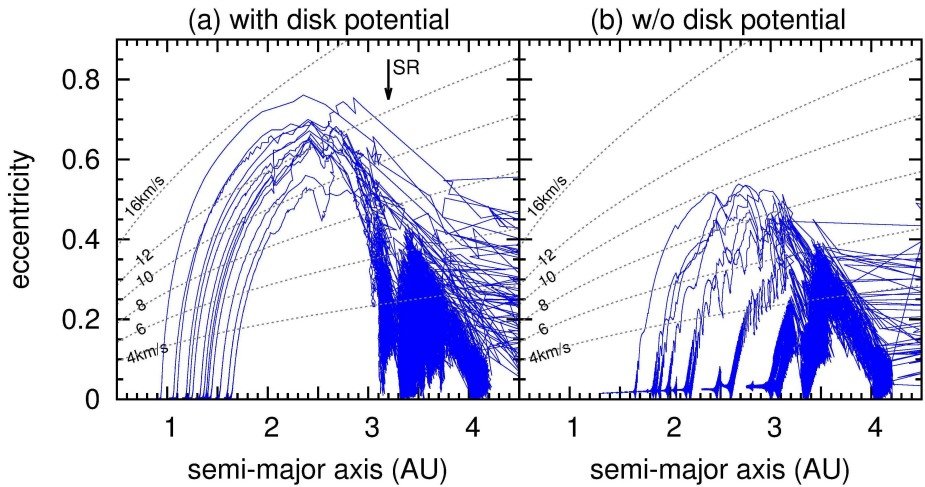}
\caption{Evolution of $a$ versus $e$ for a 300 km planetesimal starting 
from $a=4.1$ AU. Left: with disk potential. The location of the secular resonance is indicated by an arrow.  Right: without disk potential.
\label{fig:wwodisk}}\end{center}
\end{figure*}

Chondrule formation induced by shock waves requires the relative velocity to be on the order of $\ga 7~{\rm kms^{-1} }$ for a partial melt of submillimeter-sized dusts in a gas disk with a density of $\rho \sim 10^{-9}$ ${\rm g cm^{-3}}$ (e.g., Hood 1998; Iida et al. 2001; Desch \& Connolly 2002). 
Although the maximum velocities of planetesimals obtained in the previous simulations of planetesimal evolution in resonances suggest that chondrule formation by bow shocks is likely, the highest speeds obtained ($\la 8 $ ${\rm km s^{-1}}$) are rare and only marginally achieve efficient formation. 
Furthermore, in the asteroid belt of the minimum-mass disk 
($\rho \sim 10^{-10}$ ${\rm g cm^{-3}}$) where the resonances exist, a larger relative velocity $\ga 10$ ${\rm km s^{-1}}$ would be preferable to ensure complete melting of the 1 mm-sized dust.  
If the planetesimals can achieve a relative velocity higher than 
$10$ ${\rm km s^{-1}}$ more frequently during orbital migration, the ubiquitous existence of chondrules could be explained more satisfactorily.

In previous works regarding planetesimals in resonances, the gravity of the gas disk and planets other than Jupiter was neglected, i.e., the effect of secular resonances was neglected.
However, as we described in the previous paragraph, the effective shock-heating of chondrules requires a relatively dense gas disk, at least on the order of that of the minimum-mass disk.
Such a gas disk provides not only the drag force but also the gravitational force and causes secular resonance.
The gravitational potential of the disk precesses the Jovian pericenter.
When the precession rate coincides with the precession rate of the planetesimals, a secular resonance arises, which enhances the eccentricities of the planetesimals.  
Such a secular resonance occurs between 2~AU and 4~AU in a disk of density 
$\sim$ 0.1--5 times that of the minimum-mass disk (e.g., Heppenheimer 1980; Lecar \& Franklin 1997; Nagasawa, Tanaka, \& Ida 2000; Nagasawa, Ida, \& Tanaka 2001, 2002). 
Even if the resonance is {\it not sweeping}, its existence causes a high-amplitude oscillation of the eccentricity and further excitation of the relative velocity in the vicinity of the secular resonance.
In chondrule formation induced by planetesimal bow shocks caused by Jovian perturbation, secular resonance inevitably occurs and plays an important role.

Planetesimal bow shocks may also contribute to the origin of the crystalline silicate in comets.  
The presence of crystalline silicate in comets has been confirm though infrared observations of dust grains in a number of cases (Bregman et al. 1987; Molster et al. 1999; Hanner \& Bradley 2004). 
It is thought that crystalline silicate is formed in the protoplanetary disk because the silicate dust in the interstellar medium is almost entirely amorphous (Li, Zhao, \& Li 2007).
Experimental studies on the thermal annealing of amorphous silicate show that the formation of crystalline silicates requires temperatures above 800~K (Hallenbeck, Nuth, \& Daukantas  1998).  
In contrast, the composition of the gas in cometary comae indicates the preservation of interstellar ice in the cold outer nebula (Biermann, Giguere, \& Huebner 1982). 
It is unclear why these two materials, which have contradicting heating records, co-exist in comets (see Yamamoto \& Chigai 2005; Tanaka, Yamamoto, \& Kimura 2010; Yamamoto et al. 2010). 
It is possible that the amorphous silicates crystallize through shock heating. As shown later, a number of planetesimals are likely scattered by the Jupiter resonances. This mechanism may explain the incorporation of both high- and low-temperature materials in comets.

In this letter, we study the evolution of planetesimals in the gas disk, including the effect of secular resonances caused by the gas disk potential, and determine whether the relative velocity is sufficient for chondrule formation.  
In Section \ref{sec:results}, we briefly describe the setup of the numerical simulations and present results. 
We show that secular resonance excites most of the planetesimals up to $e \ga 0.6$  ($v_{\rm rel} \ga 12$ $ {\rm km s^{-1}})$. 
Finally, we summarize and discuss our results in Section \ref{sec:conclusions}.

\section{NUMERICAL RESULTS}
\label{sec:results}

We investigate the orbital evolution of test particles perturbed by Jupiter and the disk using the {time-symmetric fourth-order Hermite code which has an advantage in the precise long-term calculations of pericenter evolution and detection of close encounters (e.g., Kokubo \& Makino 2004 and references therein). }
The protoplanetary disk provides a background gravitational field that induces secular resonance and gas drag.
We use a thin disk potential for the minimum-mass disk as described by Ward (1981).
The gap in the disk where Jupiter exists is neglected.
We perform simulations both with and without the disk potential and compare the results. 
When we include the disk potential, the change in rotational speed of the gas disk due to its own gravity is taken into account. 
We assume planetesimals 300~km in size with a material density of $\rho_{\rm mat} =3$ ${\rm g cm^{-3}}$ to calculate the gas drag.
{From the bow shock simulations of chondrule formation, it was shown that larger planetesimals ($\ga$ 1000 km) is required to account for the thermal history, particularly cooling rates (Ciesla, Hood, \& Weidenschilling 2004; Boley, Morris, \& Desch. 2013). It is shown by previous studies of the planetesimal evolution in the gas disk that the larger planetesimals obtain larger eccentricities because of the weakened gas drag. In this letter, we select 300~km planetesimals to show the required relative velocity for the chondrule melting is achieved under the gas drag. We tested other sizes (100~km, 300~km, 500~km, and 1000~km) and confirmed that our conclusion hardly changes.}

\begin{figure}
\includegraphics[width=.16\textwidth, bb=0 0 130 300]{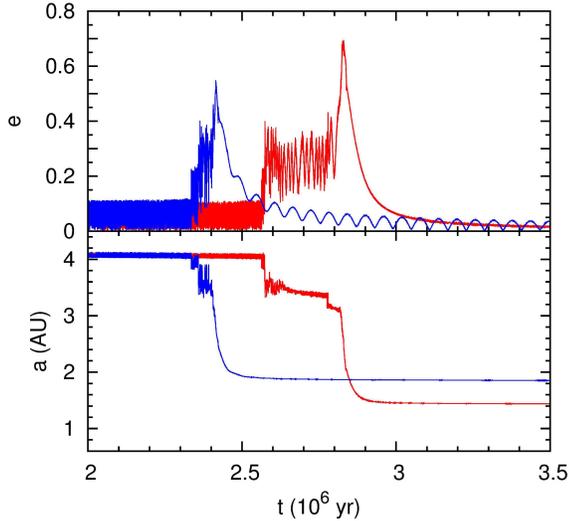}
\caption{Evolution of the semi-major axis and eccentricity. 
Red and blue lines show the simulation with 
and without the disk potential, respectively.
\label{fig:tae}}
\end{figure}

We adopt the gas drag force given by Adachi, Hayashi, \& Nakazawa (1976). 
The specific characteristics of the gas drag have little effect on the maximum eccentricity of the planetesimals (Marzari \& Weidenschilling 2002).
We set the drag coefficient, which varies with the Mach number and the Reynolds number (Tanigawa et al. in prep), but its choice is not essential in our simulations.

We use the current size, eccentricity, and semi-major axis of Jupiter, which are $1M_{\rm J}$, $e=0.048$ and $a=5.2$~AU, respectively. 
With such parameters, the secular resonance caused by the minimum-mass disk occurs at around 3.2~AU.  
We start our simulations putting planetesimals at 4.1~AU, which is just outside the 3:2 resonance region.
{When we start simulations from inside of 3AU, the eccentricity is kept smaller than 0.2 in the case of planetesimals larger than 100 km in size regardless of whether the disk potential is included.
Accordingly, the planetesimals hardly migrate since their migration timescales exceed the life-time of gas disk.}

The trajectories of 20 planetesimals 300~km in size with different initial orbital angles are plotted in Figure \ref{fig:wwodisk}. 
The evolutions in which the disk's self-gravity is included are shown in the left panel and the evolutions without the disk potential are shown in the right panel.

The planetesimals migrate inwards due to the gas drag. 
As already shown in previous works, the resonances between 3~AU and 4~AU increase the eccentricities of migrating planetesimals. 
Since the 3:1 resonance at $\sim$ 2.5~AU is separated from the other resonances, the eccentricity is damped before the 3:1 resonance is reached as shown in the right panel. 
On the other hand, in the left panel, the increase in eccentricity continues until the 3:1 resonance due to the extra excitation caused by the secular resonance.

The typical velocity of the planetesimals relative to the gas disk can be estimated using $v_{\rm rel} \sim e V_{\rm Kep}$, where $V_{\rm Kep}$ is the Kepler velocity at that semimajor axis $a$ and $e$ is its eccentricity (Adachi et al. 1976).  
{ Note that $e V_{\rm Kep}$ is the redial velocity at the location of $r=a$ (where $r$ is the distance from the star). 
In the case of low eccentricity, the relative velocity $e V_{\rm Kep}$ corresponds to the maximum value and the smallest relative velocity (1/2 $e V_{\rm Kep}$) occurs at its pericenter and apocenter. 
In high eccentricity cases ($e \ga 0.5$), the actual maximum relative velocity is achieved between $r=a$ location and the pericenter and its magnitude is enlarged from $e V_{\rm Kep}$.}
In Figure \ref{fig:wwodisk}, the relative velocity ($e V_{\rm Kep}$) is shown by dotted gray lines.  When the disk potential is not considered, the maximum speed rarely exceeds 10~${\rm  km s^{-1}}$; when it is considered, however, the maximum speed exceeds 10~${\rm km s^{-1}}$ for all planetesimals.

Figure \ref{fig:tae} shows typical evolutions of the semi-major axes and eccentricities for two cases: including the disk potential (red lines) and excluding the disk potential (blue lines).  
The planetesimals start migration when their eccentricities are pumped up by Jupiter. 
Since the initial location of the planetesimals is near the chaotic resonance-overlapping region, the period for which the planetesimals remain near 4.1~AU depends on cases. 
When they migrate to the 2:1 mean-motion resonance ($\sim 3.3$~AU), they become trapped in it. 
In the case of Figure \ref{fig:tae}, the oscillations during 2.35 Myr--2.4 Myr (blue line) and 2.6~Myr--2.8~Myr (red line) correspond to such trapping.  
In our 20 simulations, the time trapped in the resonance tends to be longer when the disk potential is considered.  
When their eccentricities are excited to $e \sim 0.4$, the planetesimals become detached from the resonance as the discussed in previous papers (e.g., Marzari \& Weidenschilling 2002).  
If there is no secular resonance between 2~AU and 3~AU, planetesimals continue rapid migration due to the gas drag until their orbits become circular.  
If secular resonance is taken into account, however, the migration is again halted temporarily at the location of the resonance ($\sim$2.8~Myr). 
The eccentricities are excited further, but with such high eccentricities, the gas drag drives away the planetesimals from the secular resonance at around $e \sim 0.7$.
The 3:1 resonance ($a=2.5$~AU) does not play as important a role as that of the 2:1 resonance or the 3:2 resonance, but with high eccentricity, it can further increase the eccentricity by a small amount ($\Delta e \la 0.1$).  
The region where the maximum speed tends to be recorded is the location of the 3:1 resonance.

\begin{figure}
\includegraphics[width=.16\textwidth, bb=0 0 100 220]{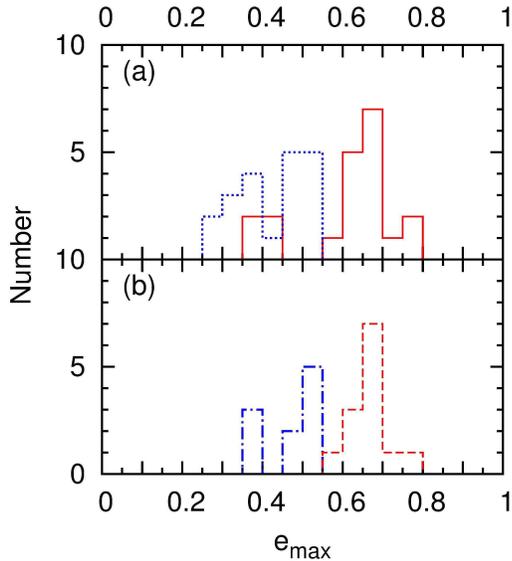}
\caption{Histogram of maximum eccentricity. Red and blue lines 
show the results for 20 simulations with and 
without the disk potential, respectively. 
Panel {\it a}: the maximum eccentricity for each planetesimal within 4.5 AU. 
Panel {\it b}: the planetesimals that experienced a Jovian encounter are omitted. 
\label{fig:hist}} 
\end{figure}

Figure \ref{fig:hist} is a histogram of the achieved eccentricities ($e_{\rm max}$) of each planetesimal during the evolution. 
The distributions obtained from the simulations with and without the disk potential are shown by red and blue lines, respectively.  
Panel~{\it a} contains all 20 simulations for each case. 
Seventeen planetesimals out of the 40 exhibit close encounters with Jupiter and enter the $a>4.5$ AU region at least one time. 
These are omitted from Panel~{\it b}.
Since the scattered planetesimals obtain $e\ga 1$ values, we count $e_{\rm max}$ when the planetesimals are within 4.5 AU. 
The groups of lower eccentricities in the bimodal distributions in Panel~{\it a} tend to correspond to the planetesimals that encountered Jupiter. 
The planetesimals remaining in the asteroid region reach high velocities due to the secular resonance, not due to the Jovian encounters.

The figure reveals that the typical maximum eccentricity with secular resonance is $\Delta e \sim 0.15$ higher than that obtained when only mean-motion resonances are considered.  
The evolution toward the 2:1 resonance basically follows a line of apocentral distance $a(1+e) \sim 4.5$ AU.
When the eccentricity excitation stops near the 2:1 resonance, $e_{\rm max}$ is $\sim 0.4$; the peak near $e_{\rm max} \sim 0.4$ (blue distribution in Figure \ref{fig:hist}) originates from this fact. 
About 1/3 of the planetesimals drop out of the 2:1 resonance in the case without the disk potential. 
On the other hand, in the case with the disk potential, all planetesimals that are not scattered by Jupiter continue on their trajectory until reaching the 3:1 resonance. 
The peak at $e_{\rm max} \sim 0.7$ corresponds to this state.  
At the location of the 3:1 resonance, $V_{\rm Kep} \sim 30 \ {\rm km s^{-1}}
(a {\rm /AU})^{-1/2} \sim 20~ {\rm km s^{-1}}$.
Thus, the typical maximum of the relative velocity is approximately  given by 
$20 \times e \ {\rm km s^{-1}}$. 
The difference between the two cases is 
$\sim 3$ ${\rm km s^{-1}}$ ($\Delta e \sim 0.15$).

{ Unlike the eccentricities, inclinations ($i$) remain lower than $10^{\circ}$ in most cases. 
It is because a strong secular resonance of Jupiter which excites the inclination ($\nu_{15}$) hardly occurs in the gas disk within Jovian orbit (Nagasawa et al. 2000, 2001, 2002).
Although planetesimals with $i=10^{\circ}$ go out of the gas disk, they stay longer than 1/5 of their orbital period within one scale height of the disk, where the gas density is comparable to that at the disk midplane.
}

\begin{figure}\epsscale{0.8}
\plotone{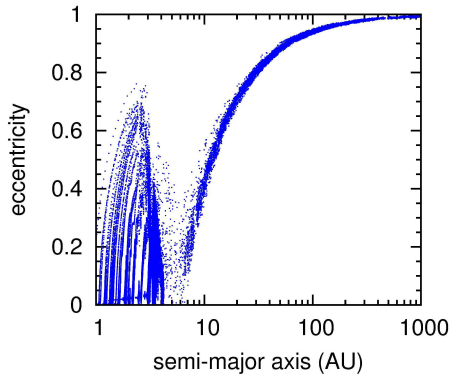}
\caption{Wide range $e$-$a$ diagram.
\label{fig:scatter}}
\end{figure}

The planetesimals beyond the $a(1+e)\sim 4.5$ AU line enter into a region of resonance-overlapping and they are scattered by close encounters with Jupiter. 
About half of the 300~km-sized planetesimals are scattered in both cases. 
This fraction would be lower for smaller planetesimals due to the stronger gas drag.  
One out of 15 scattered planetesimals returns to the region of $a<4.5$~AU, but the majority eventually reach $e \ga 1$ and leave orbit (Fig.~\ref{fig:scatter}).

\section{DISCUSSION AND CONCLUSIONS}
\label{sec:conclusions}

We studied the evolution of planetesimals in the gas disk, including the effect of secular resonance caused by the gas-disk potential.  
We found that the planetesimals attain $e\sim 0.6$ with the help of this secular resonance.  
The relative velocity of the planetesimals exceeds 12~${\rm km s^{-1}}$ around the 3:1 mean-motion resonance.  
The high-eccentricity region is restricted to a relatively narrow band of 1.5 AU--4 AU. 
In previous studies on the planetesimal evolution in the gas disk, the maximum velocity was found to be $\sim 8$ ${\rm km s^{-1}}$ ($e \sim 0.6$ at 2~AU) and such supersonic speeds were not frequent events (Marzari \& Weidenschilling 2002).  
In our simulations with secular resonance, however, about the half of planetesimals reached $e \sim 0.6$.

Our results are supportive of the possibility of chondrule formation induced by planetesimal shock waves due to Jovian resonance.  
The minimum relative speed required for melting 1 mm-sized dust at 1 AU is considered to be $\sim 7$ ${\rm kms^{-1}}$ in the minimum-mass disk (e.g., Hood 1998; Iida et al. 2001; Desch \& Connolly 2002).  
The typical $e_{\rm max}$ of $\sim 0.65$ around 3~AU to 2~AU corresponds to a velocity of $11~{\rm km s^{-1}}$--$14~{\rm km s^{-1}}$ relative to chondrule precursors rotating with the disk. 
The density of this region is $2 \times 10^{-10} {\rm g cm^{-3}}$--$7 \times 10^ {-11} {\rm g cm^{-3}}$.
With this disk gas density, the formation of 0.1 mm chondrules by shock waves requires $\sim 10$ ${\rm km s^{-1}}$--18 ${\rm km s^{-1}}$ (Iida et al. 2001), while $\sim 7 $ ${\rm km s^{-1}}$ is sufficient in 10 times denser regions. 
If the speed exceeds $20$ ${\rm km s^{-1}}$, even a 1 cm precursor evaporates, but such speeds are not realized.  
Note that the highly supersonic situation ($\ga$ 12$ {\rm km s^{-1}}$) is restricted to the 1~AU--3~AU region within the stable region of $a \ga 4.5 $AU.  
That would suggest that the chemical or taxonomic evolution of planetesimals may depend on the semi-major axis via the gas-disk density and the maximum heating caused by bow shocks.

Our results relate to the origin of crystalline silicates observed in a number of comets.  
In our simulations, about half of the planetesimals attain $e \ga 1$ and move toward the outer region of the disk due to Jovian resonances.
It was shown that such eccentric icy planetesimals with core-mantle structures are changed to rocky planetesimals due to the efficient evaporation of the icy mantle caused by shock heating, even in the region outside the snow line (Tanaka et al. 2013).
As long as the dry planetesimals are transported in the cometary region, there is no conflict with the fact that cold interstellar ice is preserved in the comae.
On the way to the outer region, the planetesimals are expected to accrete the silicate dust processed by the shock waves.
This process can explain the presence of crystalline silicates in comets, if the scattered planetesimals containing crystalline silicates become mixed with icy outer planetesimals.

In this letter, we considered 300 km-sized planetesimals in the minimum-mass gas disk. 
If we considered smaller planetesimals ($\la 100$ km), the maximum velocities would have been smaller as a result of stronger gas drag. 
On the other hand, the maximum velocity did not differ greatly to that for larger planetesimals ($\ga 100$ km); however, the ejection frequency would be enhanced due to weakened gas drag in such a case.  
The maximum speed of planetesimals depends on their mass, the eccentricity, and the semi-major axis of Jupiter through the strength of the secular resonance.
The effect of the secular resonance starts to come into play when Jupiter exceeds $\sim 1/3$ of its current mass, and the effect continues until the disk is dissipated ($\sim$ 90\%). 
Surveys changing these parameters will be conducted in subsequent work.

\acknowledgments
We are thankful for a careful and helpful review by an anonymous referee. 
MN was supported in part by JSPS KAKENHI Grant Number 25610133. 
KKT was supported in part by JSPS KAKENHI Grant Number 26108503 and 2540054.
This study was partly supported by the Grant for Joint Research Program of the Institute of Low Temperature Science, Hokkaido University.


\end{document}